\documentclass[10pt,conference]{IEEEtran}
\usepackage{hyperref}
\usepackage[most]{tcolorbox}
\usepackage{placeins}
\usepackage{tabularx}
\usepackage{booktabs}

\newtcolorbox{rqbox}{
  colback=gray!15,
  colframe=black,
  boxrule=0.8pt,
  arc=6pt,
  left=6pt,
  right=6pt,
  top=6pt,
  bottom=6pt
}
\usepackage{graphicx}
\usepackage{listings}
\usepackage{xcolor}
\usepackage{soul}
\usepackage{enumitem}
\usepackage[nameinlink,noabbrev]{cleveref}
\graphicspath{{./images/}}

\title{ADAPT: Agentic Diff-Aware Patch Testing for Pull Requests}

\author{
\IEEEauthorblockN{Vahid Haratian, Atakan Akar, and Eray T\"uz\"un}
\IEEEauthorblockA{
Bilkent University\\
Ankara, Turkey
}
}

\newtcolorbox{findingbox}{
    colback=gray!15,
    colframe=gray!60,
    boxrule=0.8pt,
    arc=6pt,
    left=6pt,
    right=6pt,
    top=6pt,
    bottom=6pt
}

\begin{document}
\bstctlcite{IEEEexample:BSTcontrol}

\maketitle

\begin{abstract}
Modern software development increasingly relies on pull requests (PRs), where developers introduce small, targeted code changes.
Once merged, these changes should be protected by focused tests that exercise the specific behavior the PR introduces.
In practice, however, developers often submit PRs without such tests, leaving the new behavior unguarded against future regressions.
Most of the techniques do not close this gap: they target entire software units and optimize objectives such as code coverage, rather than the behavior a change introduces.
Also, prior work on diff-aware testing has achieved only limited success in generating tests that capture this behavior.

We propose ADAPT (\textbf{A}gentic \textbf{D}iff-\textbf{A}ware \textbf{P}atch \textbf{T}esting), a diff-aware regression test generation approach that introduces an autonomous code generation agent to produce change-focused tests for PRs.
Given a PR, the agent analyzes the diff between the base and head versions to identify the modified functions and their context, then plans and generates candidate unit tests for the modified functionality.
To direct these tests toward the intended change, ADAPT uses a dedicated fail-to-pass (F2P) verification tool that runs each candidate test on both the base and head versions and, after F2P success, reports which executable changed lines remain uncovered; guided by this feedback, the agent iteratively refines its tests.

We evaluate ADAPT on the same 426 real-world PRs used by the prior state-of-the-art baseline, with GPT-4o and GPT-5-mini.
ADAPT+GPT-4o reaches an F2P rate of 62.0\% (264/426), compared with BLAST's reported 35.4\% (151/426), an absolute improvement of 26.6 percentage points, while ADAPT+GPT-5-mini reaches 89.9\% (383/426).
On 137 PRs shared with Change And Cover (ChaCo), ADAPT also achieves full patch coverage on 83.9\% of PRs, compared with 79.6\% for ChaCo with considerably lower time and cost.
A manual evaluation confirms that the F2P-valid tests target each PR's intended behavioral change, reaching mean scores of up to 4.54/5 for importance, 4.35/5 for assertion strength, and 4.49/5 for readability, and often covering behavior beyond the developer-written tests.

\end{abstract}


\section{Introduction}
\label{sec:introduction}

For a long time, work on automated test generation has mainly focused on creating extensive test suites for entire program units, including classes or modules~\cite{lukasczyk2022pynguin, Fraser_Arcuri_2012, holmes2020, shamshiri2015}. 
These approaches are typically assessed using holistic measures, including statement and branch coverage as well as mutation-based metrics \cite{wholeTest, shamshiri2015}. 
However, today’s software rarely changes in large, sweeping updates. 
Instead, it grows through Pull Requests (PR), each introducing small, localized changes in the codebase~\cite{Zhang2023}.
Consequently, industrial best practices also encourage developers to include focused unit tests in their PRs that directly target the newly introduced changes, reducing the risk of regression and supporting long-term code quality ~\cite{nasa2025regression, microsoft2024engineering,chromiumTesting, gitlabFrontendTesting, googleCodeReviewTests}. 
Moreover, prior work finds that practitioners expect tests to accompany any new or modified behavior; in particular, Kochhar et al.\ ~\cite{kochhar2019practitioner} surveyed 261 developers and reported an average Likert agreement rating of 4.40/5 for the principle that a bug fix should be accompanied by corresponding tests that cover the changes and prevent future regression.
In practice, however, this expectation is often unmet, and a large fraction of PRs are merged without any accompanying tests, leaving the introduced changes unprotected against future regressions \cite{beyondaccuracy2024, gousios2014exploratory, milanese2026humanagent, amore2026documentation, dipongkor2026coverage}. For example, Amore et al.~\cite{amore2026documentation} analyzed merged 
PRs across 160 popular GitHub repositories 
from January 1 to December 1, 2025, and reported a mean Test Engagement Ratio of $0.37$, meaning that on average $63\%$ of merged pull requests that modified production code did not also modify any test file. 
In this setting, achieving wide file-level coverage is no longer the main objective. Instead, the emphasis shifts toward crafting tests that deliberately target the specific change set, whether introduced through a commit or a PR, capture the new functionality, and distinguish the original behavior of the codebase from the newly introduced functionality~\cite{MaWei2020}. This change-focused view motivates Fail-to-Pass (F2P) tests ~\cite{jimenez2024swebench}, which are designed to fail on the base version of the code before the PR changes and pass only after the modifications are applied. Accordingly, F2P has become more common in recent work on change-focused test generation as a criterion for assessing whether generated tests capture the behavioral changes introduced by patches~\cite{kitsios2025automated, wang2026icore}.

BLAST combines LLM-based generation and SBST to produce F2P tests from an issue description and its resolving patch~\cite{kitsios2025automated}.
It was evaluated on 426 PRs from 12 Python repositories in SWE-bench Verified~\cite{openai2024swebenchverified} and generated change-aware tests for 35.4\% of cases (151/426).
Because its predefined pipeline has limited ability to revise intermediate plans, patch-level generation may benefit from an agentic workflow that can inspect the repository, execute candidates on both versions, analyze failures, and adapt from feedback.
ADAPT adopts this workflow.

Unlike BLAST, which takes the issue description together with the resolving patch as input, ADAPT generates tests directly from the code change alone, without relying on any issue description~\cite{liu2019automatic}.
ADAPT is structured around a diff-aware task formulation that provides the agent with PR-specific context, including the base and head versions of the code, together with an iterative verification workflow that evaluates and refines candidate tests so they will aim for the change introduced by the PR. 
To ensure that generated tests encode the semantics of the change, we incorporate an F2P verification tool that checks whether a candidate test fails on the base version and passes on the head version, and provides feedback that helps the agent iteratively refine candidate tests toward generating unit tests that specifically target and validate the behavioral differences introduced by the PR.

We evaluate ADAPT against BLAST on the same 426-PR SWE-bench Verified subset using the same F2P metric and GPT-4o, and additionally evaluate GPT-5 mini~\cite{openai2024swebenchverified,kitsios2025automated}.
For RQ4, we compare ADAPT and Change And Cover (ChaCo) with GPT-5 mini on 137 shared PRs from pandas, SciPy, and Qiskit, measuring full patch coverage~\cite{zhou2024change}.

\newcommand{\RQone}{Does ADAPT outperform BLAST in generating F2P test cases for PRs?}
\newcommand{\RQtwo}{Does the F2P verification tool act as an active refinement signal during generation rather than a final check?}
\newcommand{\RQthree}{Do F2P-valid tests generated by ADAPT meaningfully target the behavioral changes introduced by the PR?}
\newcommand{\RQfour}{Does ADAPT outperform ChaCo in generating change-focused test cases with full patch coverage for PRs?}

 \textbf{RQ1:} \RQone

\textbf{RQ2:} \RQtwo

 \textbf{RQ3:} \RQthree

\textbf{RQ4:} \RQfour

Our contributions are: (1) ADAPT, a diff-aware agentic generator of change-focused PR tests; (2) an F2P verifier that drives refinement and reports uncovered executable changed lines; (3) a controlled comparison with BLAST on 426 PRs and an analysis of verifier feedback; (4) a manual evaluation of F2P-valid tests; and (5) a 137-PR comparison with ChaCo covering patch coverage, time, and cost.

Our replication package provides the data, execution evidence, and complete ADAPT trajectories.\footnote{\url{https://figshare.com/s/18158ef3257f31b8598a}}

\section{Background and Related Work}

\paragraph{Search-Based Software Testing}
SBST uses metaheuristic search to optimize structural objectives: EvoSuite targets branch, line, and mutation coverage in Java~\cite{Fraser_Arcuri_2012}, while Pynguin applies similar techniques in Python~\cite{lukasczyk2022pynguin,lukasczyk2022empirical,bhatia2024}.
Evolutionary tests can be difficult to maintain because they often contain nonintuitive inputs~\cite{palomba2016quality,grano2018readability,palomba2016diffusion}.
Moreover, whole-module structural coverage does not ensure that a test captures the behavioral change introduced by a PR~\cite{haratian2026praware,serra2019tenyears}.
ADAPT instead uses diff-aware generation to exercise PR-modified code.

\paragraph{LLM-Based Unit Test Generation}
LLMs have been widely studied as replacements or complements to search-based test generation~\cite{tufano2021unit,schafer2023empirical,yuan2024nomoremanual,pizzorno2025coverup,ryan2024codeaware,deljouyi2024utgen,hossain2024togll,alshahwan2024automated,Dakhel2024,wang2024hits}.
Prior work spans synthesis from source and focal context~\cite{tufano2021unit}, empirical evaluation of executability and coverage~\cite{schafer2023empirical}, iterative feedback and repair~\cite{yuan2024nomoremanual,konstantinou2025yate}, coverage- and code-aware prompting~\cite{pizzorno2025coverup,ryan2024codeaware}, and improvements to readability and test oracles~\cite{deljouyi2024utgen,hossain2024togll}.
Industrial deployment additionally requires build-and-run gates and controls for flakiness and redundancy~\cite{alshahwan2024automated}.

These execution-feedback and quality improvements~\cite{yuan2024nomoremanual,ChatUniTest2024} increasingly rival traditional SBST~\cite{straubinger2025mutationtestingiterativelarge,bhatia2024}.

\paragraph{Change-Focused Test Generation for PRs and Commits}
WAFLGo is a directed greybox fuzzer that targets commit-related regressions by biasing exploration toward modified and data-dependent code~\cite{waflgo2024}.
Cleverest repeatedly queries an LLM to synthesize inputs that expose pre- and post-change behavioral differences from a commit message or diff~\cite{cleverest2025}.
Both produce inputs rather than executable, source-integrated unit tests.

BLAST instead combines SBST and LLM prompting with issue, patch, focal, and test context, using SBST-generated passing tests as grounding for executable F2P tests~\cite{kitsios2025automated}.

Testora flags generated behavior that conflicts with documented PR intent, whereas ADAPT preserves the intended patch behavior~\cite{pradel2026testora}.
FuzzSight uses fuzzing to support code review~\cite{charoenwet2025fuzzsight}.
SWT-Bench and e-Otter++ generate issue-reproducing tests from reports and pre-fix code before a patch exists~\cite{swt-bench,ahmed2026heterogeneous,wang2026icore}.
ADAPT and BLAST instead start from an implemented PR and generate regression guards for its introduced behavior~\cite{kitsios2025automated}.
Related work also updates existing tests after code changes~\cite{hu2023ceprot,zhang2025testupdater}.

Closest to our setting, ChaCo augments human-written tests to improve changed-region coverage~\cite{zhou2024change}.
ADAPT instead generates tests directly from the PR diff and first requires F2P, using uncovered executable changed lines only as post-F2P refinement feedback.

\paragraph{Agentic LLM Workflows for Software Engineering}
Agentic software engineering tools include Claude Code~\cite{anthropic2024claudecode}, Cursor~\cite{cursor2024}, OpenHands~\cite{wang2024openhands}, and SWE-agent~\cite{yang2024sweagent}.
They combine LLM reasoning, tool use, execution feedback, and iterative refinement to navigate and modify multi-file repositories~\cite{RepairAgent2025,autocoderover2024,jiang2026agenticsurvey}.
Existing academic work has focused primarily on file-level or standalone unit-test generation~\cite{jain2025testforge}; to our knowledge, no prior work proposes agentic PR-level test generation.
Built on OpenHands~\cite{wang2024openhands}, ADAPT fills this gap by refining diff-aware tests through base/head F2P feedback.

\section{Methodology}
\label{sec:methodology}

\begin{figure}[t]
    \centering
    \includegraphics[width=\columnwidth]{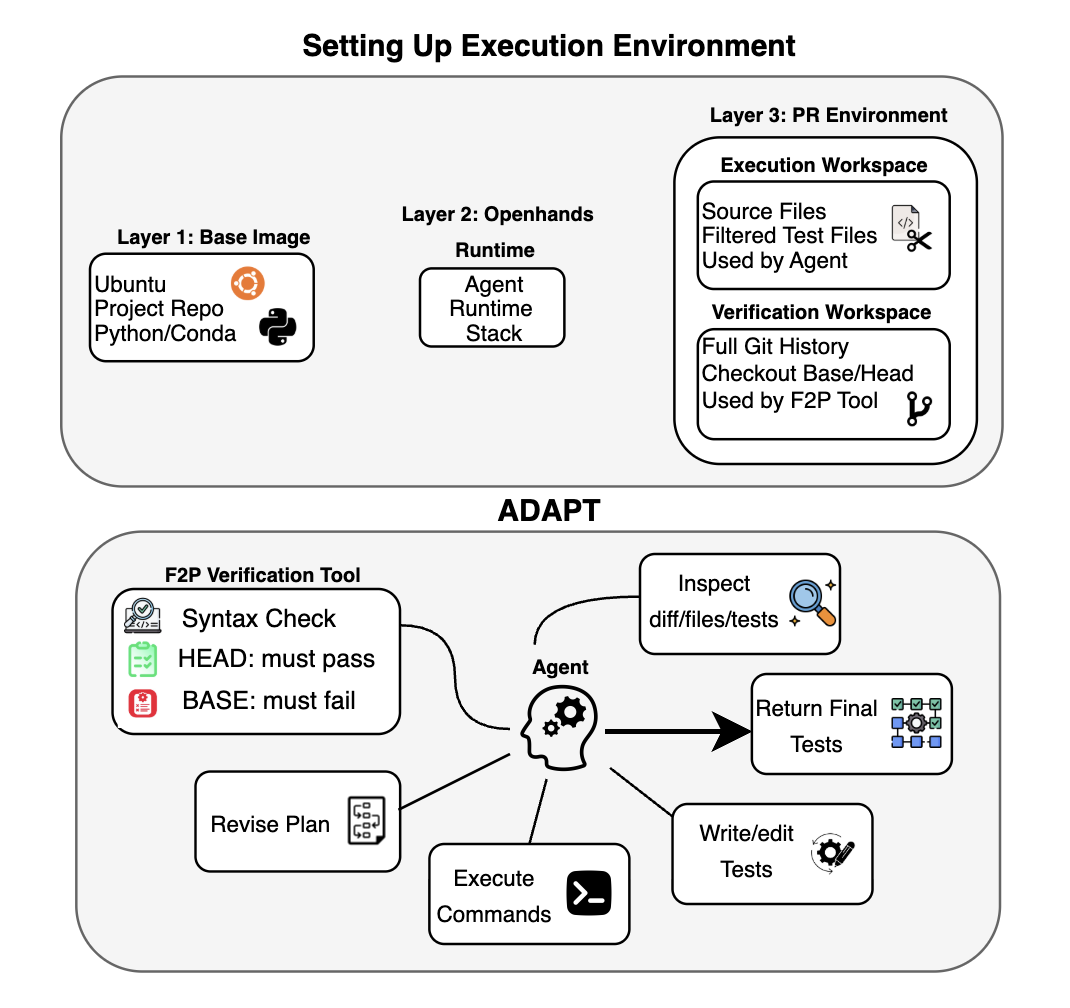}
    \caption{Overview of the ADAPT framework, including PR context extraction, execution environment setup, agent-driven test generation.}
    \label{fig:adapt-overview}
\end{figure}

In this section, we present the methodology of our study and the structure of ADAPT. 
Figure~\ref{fig:adapt-overview} provides an overview of the entire evaluation pipeline of ADAPT.

\subsection{Dataset}
To enable a direct and fair comparison with our baseline, BLAST~\cite{kitsios2025automated}, we evaluate ADAPT on the same dataset.
This dataset is a 426-instance subset of SWE-Bench Verified~\cite{openai2024swebenchverified, jimenez2024swebench}, and its instances are drawn from 12 popular open-source Python repositories, including Django~\cite{django}, SymPy~\cite{sympy2017}, Matplotlib~\cite{hunter2007matplotlib}, and Scikit-learn~\cite{pedregosa2011scikit}.
Each instance is built around a real-world issue and the PR that resolves it, together with a Docker environment in which the correctness of the changes can be verified.
For each instance, SWE-Bench provides two patches extracted from the PR, a gold patch and a test patch.
The gold patch contains the source-code changes that fix the issue, while the test patch contains the test cases added or modified to verify that the fix behaves as intended.
As a result, every instance reflects a realistic software-development scenario in which a behavioral change is introduced through a PR.
Because the dataset is the verified subset of SWE-Bench, each PR was manually checked by the dataset curators during its creation.
These tests are imperfect references for PR behavior~\cite{openai2026sweverified, shi2026promax, yu2026sweabs}; ADAPT's F2P verdicts do not rely on them.

\subsection{ADAPT}
\label{sec:ADAPT}

ADAPT is an autonomous test-generation agent built on top of OpenHands~\cite{wang2024openhands}.
ADAPT works directly from the PR information already provided by SWE-Bench Verified, namely the golden patch (the source-code changes) and the test patch that identifies the affected test files.
OpenHands provides the interactive action--observation interface through which ADAPT can inspect repository files, execute shell commands, write tests, and iteratively refine them based on execution feedback.
On top of OpenHands, ADAPT adds task-specific constraints and a custom F2P verification tool that reports whether a generated test captures the behavioral change introduced by the PR and guides refinement toward the semantic difference between the base and head versions of the repository.

Unlike issue-reproduction approaches, ADAPT does not take the issue description as input and does not attempt to reproduce a reported bug.
Instead, ADAPT generates test cases directly from a given source-code change, treating the modified implementation as the specification of the behavior to be tested.
Rather than following a test-first style, ADAPT writes tests after a change is made to capture the new behavior and guard against future regressions.
This reflects common practice, as TDD sees challenges in industrial adoption and writing tests after the implementation remains a widespread need~\cite{causevic2011factors, tosun2017industry, StaegemannTDD}.

\subsubsection{Task Definition and Constraints}
At the beginning of each session, the agent receives a single structured prompt that defines the task and provides the PR context.
The prompt specifies two linked objectives: the generated tests must satisfy the F2P criterion---pass on the head version of the repository while failing on the base version---and the agent's F2P tests must together execute every executable changed line of the PR diff (100\% patch coverage).
F2P alone is not treated as success in the prompt: a test can be valid F2P while exercising only a small fraction of the changed lines.
It also explicitly instructs the agent not to modify any existing source or test files, and to add its tests only to the new test files that ADAPT provides for this purpose.
This restriction prevents the agent from repairing the code in order to satisfy the test objective, and ensures that the generated test must expose the behavioral difference introduced by the PR.
The prompt further outlines a high-level workflow for the agent.
The agent first inspects the change and reads the modified implementation files to understand the new behavior.
It then examines the existing test files to learn project-specific testing conventions.
Based on this, it designs candidate tests that target the changed behavior and writes them to the designated files.
Finally, it executes these tests and iteratively refines them using the feedback returned by the F2P verification tool, including the patch-coverage report described below when the tests are already F2P-valid.
The complete prompt is available in the replication package.

To keep the agent's edits separate from the existing code base, ADAPT does not allow the agent to write into the repository's original test files.
Instead, ADAPT inspects the files touched by the PR's test patch and, for each of them, creates a dedicated ADAPT-equivalent test file.
The names of the files touched by the test patch are included in the prompt, so the agent can open them, learn the relevant testing conventions, and decide how to structure its own tests.
The agent then writes its tests only into these ADAPT-equivalent files, while the original test files remain unchanged.
The F2P verification tool considers only these newly created files and ignores every other file in the repository, which keeps verification focused on the agent's own tests.

\subsubsection{Generation Workspace}
During generation, the agent operates inside a repository workspace containing the code base of the repository where the PR changes are applied, we call this Testbed.
This lets the agent ground its reasoning in the actual repository state instead of relying only on the summarized prompt.
To avoid solution leakage we make sure that the test patch provided in the PR and the git history are not available in this workspace. 
While exploring this workspace, the agent relies on the repository's existing tests as a reference, and these reference tests play two roles in the process.
First, they help the agent learn repository-specific testing conventions such as import patterns, fixture usage, assertion style, and test class structure.
Second, ADAPT constrains the agent to reuse only imports that already appear in these reference tests, which reduces hallucinated import paths and helps keep the generated tests compatible with the project's test infrastructure.

The agent's interaction with the environment follows the standard OpenHands~\cite{wang2024openhands} action--observation loop.
At each step, the model selects an action such as reading a file, executing a shell command, or writing a test file, and the environment returns the resulting observation.
These observations include command output, runtime errors, and test-runner feedback, and they are appended to the interaction history so that subsequent decisions can be conditioned on previous outcomes.
This is the mechanism through which the agent performs incremental reasoning and self-correction.
For example, if a generated test contains a syntax error, import error, or failing assertion on the current repository state, OpenHands returns the raw terminal output to the model, allowing the agent to revise the test accordingly.
In this sense, OpenHands already supports iterative checking of whether a test is executable and behaves correctly on the currently checked-out version of the repository.

ADAPT adds a set of guardrails to OpenHands, both in the prompt and in its code, to keep the agent from falling into unproductive loops.
We observed that the agent would sometimes call a tool, receive an error, and then reissue the exact same call with the same input without changing anything, which led to an endless loop.
This happened most often with the file-editing tool, where the agent repeatedly issued the same edit until it exhausted its step budget and failed.
In the code, we detect these repeated identical calls and prevent the agent from issuing them.
In the prompt, we alert the agent to this failure mode and instruct it to confirm, before every tool call, that the call differs from the prior call.

\subsubsection{F2P Verification Tool}
Testbed alone, however, cannot establish the full F2P property.
The testbed exposes only the current version of the repository and does not provide usable Git history for switching between commits, so the agent cannot independently evaluate the same generated test against both the pre-change and post-change versions.
A separate mechanism is therefore required to determine whether a generated test truly satisfies the F2P property across the two versions.
The F2P verification tool fills this gap, and it is the central component that ADAPT adds.
The agent invokes the tool as part of its normal action loop, and the tool's full output is returned to the agent as an observation.

The verification tool gathers all the tests the agent has generated in the testbed where it is running.
It then moves these tests into a separate workspace that retains the repository together with its full Git history.
Within this workspace, each generated test is copied to the same path it occupies in the original repository.
The tool first switches to the head version of the repository, installs it, and runs the generated tests against it.
It then switches to the base version and runs the identical tests again.
Keeping generation and verification in two separate workspaces lets ADAPT develop tests in one place while performing the controlled base/head comparison in another.
The verification tool returns substantially richer feedback than a single binary flag.
For both the head and base runs, it reports whether each generated test compiled, whether it passed, and the raw test-runner output, so the agent can observe assertion failures, stack traces, import errors, and other runtime messages from both versions.
Using these results, the tool classifies each generated test against the F2P criterion: a test is F2P only when it fails on the base version and passes on the head version.
The tool reports success when at least one generated test satisfies this condition, and otherwise reports failure together with the non-F2P and error outcomes that the agent can act on.

The tool also distinguishes between different failure modes and exposes them to the agent in an interpretable form.
If no test passes on the head version, the agent receives a signal that its assertions are likely incorrect or that the test targets the wrong behavior.
If no test fails on the base version, the agent is told that the test is too generic and does not actually detect the change.
These two cases are highly informative for refinement~\cite{dai2025feedbackeval}.
In the first case, the agent typically re-reads the modified implementation or reruns the test on the current version to understand the correct post-fix behavior and adjust its assertions.
In the second case, it usually strengthens the assertions, changes the inputs, or targets a more specific branch or corner case that better reflects the behavioral change introduced by the PR.
Figure~\ref{fig:f2p-loop} illustrates this refinement process on a concrete instance (\texttt{sympy-20801}), showing how the two diagnostic signals successively steer the agent from a non-running test, to a too-generic test, and finally to a valid F2P test.

Once at least one generated test is F2P-valid, the same verification tool additionally measures \emph{patch coverage}.
It identifies the F2P test nodes from the dual-version run, re-executes only those tests under a coverage tracer on the head version, and intersects the executed lines with the executable changed lines of the source-only PR diff.
The tool then returns a patch-coverage block that lists, for each changed file, which changed lines the F2P tests covered and which remain uncovered, together with the source text of the uncovered lines.
Non-executable lines such as comments, blanks, and docstrings are excluded from the denominator, and only F2P tests contribute to the measurement: a non-F2P test does not increase patch coverage.
This report is feedback for the agent rather than a separate tool: the agent continues to call \texttt{/workspace/verify\_f2p.sh}, and patch coverage appears in the observation only after F2P success.
When uncovered lines remain, the prompt directs the agent to extend existing F2P tests or add new ones that drive execution through those lines, and to keep iterating until the uncovered list is empty.
Patch-coverage measurement failures are non-fatal: they do not overturn an F2P verdict.

This separation between OpenHands and F2P verification is central to ADAPT.
OpenHands provides the vanilla code generation engine to write the candidate tests. 
The F2P verification tool, in contrast, determines whether a candidate test actually distinguishes the head and base versions in the intended manner, and, after F2P success, reports how much of the changed source the F2P tests exercise.

\subsection{Experimental Setup}

\subsubsection{Execution Environment}
To ensure reproducible and isolated execution, each PR instance runs inside its own container that we build on top of the original SWE-Bench Verified docker image for that instance.
We reuse these images rather than building the environment from scratch, because each image already contains the repository checked out in the testbed, together with a Conda environment in which all dependencies required to run the tests are installed.
On top of this image, we install the OpenHands runtime so that the agent can inspect files and execute commands.
We then apply the gold patch, which brings the source code in the testbed to the head (fixed) version of the PR while leaving the developer-written tests at their base version.
This produces a hybrid repository state in which the agent works against the updated implementation but is never shown the tests added by the PR.
As an anti-cheating measure, we also remove the \texttt{.git} folder from the testbed so that the agent cannot recover the developer-written tests by traversing the commit history.
Finally, we create the \texttt{f2p\_workstation}, a separate workspace that preserves the full Git history of the repository and is used exclusively by the F2P verification tool to run generated tests against the base and head commits.
Because we build on the original SWE-Bench images, the Conda environment is already aligned with the benchmark, so no additional dependency installation is needed to run the tests.
As the final step of image construction, we run a verification script that executes the developer-written tests inside the \texttt{f2p\_workstation} to confirm that the environment is correctly configured and ready to use.
This check helped us detect and debug several cases where our pipeline was misconfigured.

\subsubsection{Models and Configuration}
We run the evaluation with two models, \texttt{gpt-4o-2024-08-06} and \texttt{gpt-5-mini-2025-08-07}, which we refer to as gpt-4o and gpt-5-mini, respectively, throughout the rest of the paper.
We include gpt-4o because it is the model used by BLAST, which allows us to compare ADAPT against the baseline under the same model.
For each PR instance, we cap the agent at a maximum of 50 steps and a maximum runtime of 30 minutes to complete the task.
For gpt-5-mini, we set the reasoning effort to high, whereas gpt-4o does not support the reasoning-effort setting.
All other parameters, such as temperature, are left at the OpenHands default values.
For the ChaCo comparison in RQ4, both ADAPT and ChaCo use GPT-5 mini.
ADAPT is capped at 100 agent steps and 60 minutes per PR, while ChaCo generates up to 20 candidate tests per PR, uses up to four feedback rounds for each candidate, and runs with three workers.

\newtcolorbox{agentbubble}{
    colback=blue!6, colframe=blue!30!black, boxrule=0.4pt,
    arc=6pt, left=7pt, right=7pt, top=4pt, bottom=4pt,
    width=\linewidth, before skip=3pt, after skip=3pt}

\newtcolorbox{actionbubble}{
    colback=teal!6, colframe=teal!40!black, boxrule=0.4pt,
    arc=6pt, left=7pt, right=7pt, top=4pt, bottom=4pt,
    width=\linewidth, before skip=3pt, after skip=3pt}

\newtcolorbox{verifierbubble}{
    colback=purple!5, colframe=purple!40!black, boxrule=0.4pt,
    arc=6pt, left=7pt, right=7pt, top=4pt, bottom=4pt,
    width=\linewidth, before skip=3pt, after skip=3pt}

\begin{figure}[t]
\centering
\begin{tcolorbox}[
    colback=gray!4, colframe=black!70, boxrule=0.5pt, arc=2pt,
    left=5pt, right=5pt, top=4pt, bottom=4pt]
\footnotesize
\textbf{Task.} In SymPy, comparing the number \texttt{0.0} (a "falsy" float) for
equality returned an incorrect result due to a missing edge-case check. This PR
adds a guard so that such values now compare correctly. The agent must write a test that fails on the original (buggy) version
and passes on the patched version, thereby capturing the change.\\[-5pt]

\textcolor{black!55}{\rule{\linewidth}{0.4pt}}\\[2pt]

\begin{agentbubble}
\textbf{\textcolor{blue!45!black}{Agent:}} \textit{``\ldots I will modify the
test to ensure \texttt{CustomNumber} is instantiated correctly\ldots Let me
update the test file.''}
\end{agentbubble}

\begin{actionbubble}
\textbf{\textcolor{teal!45!black}{Action:}} {\ttfamily \$ /workspace/verify\_f2p.sh}
\end{actionbubble}

\begin{verifierbubble}
\textbf{\textcolor{purple!55!black}{Diagnostic Feedback:}} {\ttfamily\scriptsize
NOT F2P: No tests pass on HEAD (fixed version). Fix this FIRST: your test must
PASS on HEAD before BASE is checked.}
\end{verifierbubble}

\textcolor{black!40}{$\big\downarrow$ \scriptsize the agent repairs setup/imports over several iterations until the test runs on head}\\[3pt]

\begin{agentbubble}
\textbf{\textcolor{blue!45!black}{Agent:}} \textit{``The tests are passing on
both HEAD and BASE, which means they are not valid Fail-to-Pass\ldots I will
refine the assertions to target the exact behavior of the \texttt{if not self}
check\ldots''}
\end{agentbubble}

\begin{actionbubble}
\textbf{\textcolor{teal!45!black}{Action:}} {\ttfamily \$ /workspace/verify\_f2p.sh}
\end{actionbubble}

\begin{verifierbubble}
\textbf{\textcolor{purple!55!black}{Diagnostic Feedback:}} {\ttfamily\scriptsize
NOT F2P: Some tests passed on HEAD but none also fail on BASE. Make your
assertions specific to the fixed behavior so they fail on BASE.}
\end{verifierbubble}

\textcolor{black!40}{$\big\downarrow$ \scriptsize the agent strengthens the assertion to aim for change}\\[3pt]

\begin{actionbubble}
\textbf{\textcolor{teal!45!black}{Action:}} {\ttfamily \$ /workspace/verify\_f2p.sh}
\end{actionbubble}

\begin{verifierbubble}
\textbf{\textcolor{purple!55!black}{Diagnostic Feedback:}} {\ttfamily\scriptsize
SUCCESS: 1 valid F2P test(s)!}
\end{verifierbubble}
\end{tcolorbox}
\caption{An ADAPT refinement trajectory on \texttt{sympy-20801}. The
\texttt{verify\_f2p.sh} tool returns an interpretable verdict.}
\label{fig:f2p-loop}
\end{figure}

\subsubsection{Evaluation Metric (F2P)}
\label{sec:f2p-verification}
At its core, F2P defines a test that targets the behavioral change introduced by a changeset.
Under this criterion, a generated test is considered valid if it fails when executed on the base (buggy) version of the code and passes when executed on the head version after the PR is applied.
In other words, the test must expose the incorrect behavior present in the base version of the repository while succeeding once the fix has been introduced.
The F2P criterion has become a widely used evaluation principle in recent software engineering benchmarks, particularly following its adoption in SWE-Bench~\cite{jimenez2024swebench}.

Our primary evaluation criterion for RQ1--RQ3 remains F2P.
A coverage signal that only asks whether some test executes a modified function is a weak substitute for this criterion~\cite{grano2020pizza}: the repository's existing tests often already reach those lines, so a generated test can increase structural coverage without asserting the behavior the PR introduces.
F2P answers that behavioral question directly, since a test can fail on the base version and pass on the head version only if it asserts the specific change.
For this reason, and following BLAST~\cite{kitsios2025automated}, the BLAST comparison uses F2P as the shared success metric.

Patch coverage plays a different role.
As described above, once tests are F2P-valid, ADAPT measures the fraction of executable changed source lines that those F2P tests execute and returns the uncovered lines to the agent as refinement feedback.
Only F2P tests contribute to this measurement, so patch coverage cannot replace F2P: it asks how thoroughly the already change-sensitive tests exercise the diff.
We use this signal during generation and, in RQ4, as a comparison metric against ChaCo~\cite{zhou2024change}, whose objective is closely related to covering the changed regions of a PR.

To further analyze the behavior of ADAPT, we categorize the outcome of each generation run into one of three groups.
Although the BLAST study reports results only at the level of F2P and non-F2P outcomes, we adopt this finer-grained breakdown to separate runs that complete but produce no F2P test from runs that never complete due to timeout or execution failures.
This categorization is used to provide a more detailed analysis of ADAPT's behavior, rather than for a category-wise comparison with BLAST.
For the direct comparison with BLAST, we therefore rely on the shared F2P rate as the primary comparison metric.

\begin{itemize}
    \item \textbf{F2P:} A run is classified as F2P if it produces at least one generated test that passes on the head version and fails on the base version. Failures on the base version may arise from assertion failures, exceptions triggered by incorrect behavior, or missing functionality in the pre-change code. By encoding the PR-introduced behavior as an executable test, such tests help guard against future regressions.
    \item \textbf{No F2P:} A run is classified as No F2P if it completes but none of its generated tests satisfy the F2P criterion. This typically happens when the tests execute successfully on both the base and head versions, so they do not distinguish the pre- and post-PR behavior and are uninformative for capturing the behavioral change introduced by the PR.
    \item \textbf{Error:} A run is classified as Error if it does not complete normally and therefore never yields an evaluable F2P test. This covers runs in which the agent exhausts its maximum number of steps, runs that exceed the time limit (timeout), and runs in which the model returns no trajectory at all, for example when it hits its reasoning limit on the first call and never issues a response.
\end{itemize}

\subsubsection{F2P Verification Tool Contribution}
\label{sec:rq2-method}
Because the F2P verification tool is a structural part of ADAPT rather than an optional add-on, it is necessarily invoked in every run.
The relevant question for RQ2 is therefore not whether the tool is used, but how it is used: only as a final check, or as an active signal that guides refinement during generation.
To answer this, we complement the outcome-level results with two analyses of how the agent interacts with the tool.
First, we analyze the verification-tool call trajectories.
If a candidate test satisfies the F2P criterion on the first call, we treat the tool as serving mainly as a confirmation step.
Multiple calls instead indicate that the agent used intermediate feedback to refine its tests, and when these later calls eventually reach an F2P outcome, they show that the tool provided useful guidance beyond final validation.
Second, we examine the composition of all test cases the agent submits to the tool across these calls.
For each model, we measure how many of the submitted test cases are F2P versus non-F2P.
This indicates how often the agent produces non-discriminative tests that the tool filters out, and how this efficiency changes with model capability.

\subsubsection{Manual Evaluation of F2P-Valid Tests}
\label{sec:manual-eval}
The F2P criterion is a strong indicator that a generated test captures
the behavioral change introduced by the PR. To confirm this and ensure
that F2P-valid tests are indeed high quality, we complement the
automatic check with a manual evaluation. Our goal is to verify that
these tests target a meaningful behavioral change of the PR and would
realistically guard against a future regression that reintroduced the
old behavior, rather than relying on the automatic outcome alone. We
therefore inspect a sample of F2P-valid tests directly and assess their
quality along several dimensions~\cite{grano2020pizza}. To this end, we manually evaluate a sample
of the valid tests generated by ADAPT along the questions listed in
Table~\ref{tab:rq3-questions}. The evaluation was conducted by four
annotators, two of the authors and two software engineers with five
years of software development experience each, and was performed
blind. Annotators did not know which model had
produced a given test, removing any bias toward a particular
configuration. We assess each test in isolation along three quality
dimensions rated on a five-point scale, and we additionally compare it
against the corresponding developer-written test through two
coverage-oriented yes/no questions.

\begin{table}[ht]
\centering
\caption{Questions used in the manual evaluation of F2P-valid tests.}
\label{tab:rq3-questions}
\begin{tabularx}{\columnwidth}{@{}clX@{}}
\toprule
\textbf{ID} & & \textbf{Question} \\
\midrule
\multicolumn{3}{@{}l}{\textit{Per-test quality (1--5 scale)}} \\
\addlinespace
Q1 & & \textbf{Importance:} does the test check behavior that actually
matters for the change---the intended behavior---rather than a trivial
or irrelevant case? \\
\addlinespace
Q2 & & \textbf{Assertion strength:} how precise are the checks? A weak
smoke test that only confirms the code ``runs without error'' scores
low, whereas a precise oracle that pins down the exact expected result
scores high. \\
\addlinespace
Q3 & & \textbf{Readability:} is the test clean, idiomatic, and easy to
follow---would it pass code review? \\
\midrule
\multicolumn{3}{@{}l}{\textit{Comparison with the developer-written test (yes/no)}} \\
\addlinespace
Q4 & & \textbf{Gen.\ $>$ Human:} does the generated test cover any
behavior the developer-written test does not? \\
\addlinespace
Q5 & & \textbf{Human $>$ Gen.:} does the developer-written test cover
any behavior the generated test does not? \\
\bottomrule
\end{tabularx}
\end{table}

To compare the two configurations on a common basis, we consider the
243 instances for which both GPT-4o and GPT-5 mini produced F2P-valid
tests, allowing each PR to be evaluated under both models. From this
pool we draw a uniform random sample of 70 instances, which
corresponds to a 90\% confidence level and a 10\% margin of error
under the default parameters. Each sampled test is rated independently
by all four annotators.

For every question, we aggregate the four annotators' ratings into the
following quantities. The \emph{mean} is the average rating across all
four annotators: a 1--5 score for the per-test questions and the
proportion of ``yes'' answers for the comparative questions. To capture
how consistently the annotators rated the tests, we report
\emph{Gwet's AC}, computed for every one of the six annotator pairs and
then averaged over the pairs. It is a
chance-corrected agreement coefficient that remains stable when ratings
are highly concentrated, a regime in which measures such as Cohen's
$\kappa$ can be misleadingly deflated; we use AC\textsubscript{2} with
quadratic weights for the five-point questions and AC\textsubscript{1}
for the binary yes/no questions. Finally, the \emph{mean absolute
difference} (MAD) is the average absolute gap between the ratings of
the two annotators of a pair, again averaged over the six pairs,
summarizing the magnitude of the disagreements rather than only their
frequency.

\subsubsection{Comparison with ChaCo}
\label{sec:rq4-method}
RQ4 asks whether ADAPT outperforms Change And Cover (ChaCo)~\cite{zhou2024change} in generating change-focused tests with full patch coverage.
We conduct the comparison on the 137 PRs for which both approaches have results: 44 from pandas, 49 from SciPy, and 44 from Qiskit.
Both approaches use GPT-5 mini.

ChaCo is a last-mile regression-test augmentation approach that starts from existing human-written tests and targets the changed lines those tests do not already cover.
A ChaCo run is fully covered when the union of its passing integrated generated tests covers every such residual line.
ADAPT instead generates new tests directly from the PR diff and reports full patch coverage when its F2P-valid tests together execute every executable changed source line.
Its line accounting follows ChaCo's rule of mapping each changed physical line to its owning statement, while excluding lines that the coverage tracer cannot measure.
We compare the approaches using the proportion of shared PRs for which each reaches its full-coverage objective, both overall and per repository.

We additionally compare their resource use.
Token totals are obtained by summing the prompt and completion tokens recorded in the run artifacts.
ADAPT's cost is taken from its per-run metrics, whereas ChaCo's cost is estimated from its recorded average cost per PR.
For ADAPT, elapsed time is measured from the first trajectory event through the final recorded result, with the documented harness timeout counted at its 60-minute limit.
For ChaCo, the timestamps provide the span from the first to the last generation call for each PR that produces generated tests.
The resource analysis uses the same 137 shared PRs for both approaches and reports mean and median time per PR.

\section{Results}
\label{sec:results}
In this section, we answer the four research questions introduced in the Introduction (Section~\ref{sec:introduction}).
For RQ1--RQ3, we report the results of ADAPT on the same subset of SWE-bench Verified consisting of 426 PRs used by the prior state-of-the-art baseline, BLAST \cite{kitsios2025automated}.
For RQ4, we compare ADAPT with ChaCo on their 137 shared PRs.

\subsection{RQ1: \RQone}
\label{sec:rq1-results}

\begin{table}[ht]
\centering
\caption{RQ1: Comparison of ADAPT with BLAST on 426 PRs.}
\label{tab:rq1_main}
\begin{tabular}{lrrrr}
\toprule
Approach & F2P & No F2P & Error & F2P Rate \\
\midrule
BLAST (GPT-4o)      & 151 & 275 & --  & 35.4\% \\
ADAPT (GPT-4o)      & 264 & 133 & 29  & 62.0\% \\
ADAPT (GPT-5 mini)  & 383 &   4 & 39  & 89.9\% \\
\bottomrule
\end{tabular}
\end{table}

As shown in \autoref{tab:rq1_main}, under the same-model setting (both
using GPT-4o), ADAPT achieves an F2P rate of $62.0\%$ ($264/426$),
compared with BLAST's $35.4\%$ ($151/426$)---an absolute improvement of
$26.6$ percentage points. Running ADAPT with GPT-5 mini further raises
the F2P rate to $89.9\%$ ($383/426$). Under GPT-4o, $133$ runs end
without producing any F2P test and $29$ runs end in error; under GPT-5
mini the No F2P count drops to $4$ while the error count rises to $39$. The error cases
differ in nature between the two models: for GPT-4o, $25$ of the $29$
errors are runs in which the agent exhausted its maximum number of
steps, $3$ are runs in which no trajectory was produced because the
model hit its reasoning limit on the first call and never issued a
response, and $1$ is a timeout. For GPT-5 mini, all $39$ errors are
timeouts.

Comparing dataset coverage, both models solve $243$ of the $426$ PRs and
$22$ are solved by neither. GPT-5 mini solves $140$ PRs unique to it
versus only $21$ unique to GPT-4o, largely covering GPT-4o's cases while
extending coverage.

The models also differ greatly in resource use
(\autoref{tab:rq1_resources}). GPT-5 mini consumes \$$16.62$ in total,
$29\times$ less than GPT-4o's \$$478.95$, and uses fewer tokens
($176.2$M vs.\ $297.7$M) and per PR ($413.6$K vs.\ $698.8$K),
with fewer steps per PR ($16.6$ vs.\ $22.5$). On the successfully solved
PRs the advantage is similar, with GPT-5 mini costing \$$16.15$ against
GPT-4o's \$$127.40$. Per completed run the two models take comparable
time ($196.9$s vs.\ $169.8$s) despite GPT-5 mini using fewer steps.

\begin{table}[t]
\footnotesize
\setlength{\tabcolsep}{4pt}
\caption{RQ1: Resource usage of ADAPT per model.}
\label{tab:rq1_resources}
\begin{tabularx}{\columnwidth}{@{}Xrrrrr@{}}
\toprule
Subset & Total\$ & \$/PR & Tok/PR & Steps & Time \\
\midrule
\multicolumn{6}{@{}l}{\textbf{GPT-4o}} \\
All    & 478.95 & 1.12 &  698.8 & 22.5 & 169.8 \\
Passed & 127.40 & 0.48 &  266.6 & 13.1 & 106.0 \\
\addlinespace
\multicolumn{6}{@{}l}{\textbf{GPT-5 mini}} \\
All    &  16.62 & 0.04 &  413.6 & 16.6 & 196.9 \\
Passed &  16.15 & 0.04 &  450.8 & 16.5 & 192.9 \\
\bottomrule
\end{tabularx}
\end{table}

\begin{findingbox}
\textbf{Finding 1.} Under the same model (GPT-4o), ADAPT reaches an
F2P rate of $62.0\%$ versus BLAST's $35.4\%$ ($+26.6$ points). With
GPT-5 mini, ADAPT's F2P rate rises further to $89.9\%$ at roughly
$29\times$ lower total cost (\$$16.62$ versus \$$478.95$) and fewer
steps per PR. GPT-5 mini also covers most of GPT-4o's solutions,
solving $140$ additional PRs while only $21$ are unique to GPT-4o.
\end{findingbox}

\subsection{RQ2: \RQtwo}
\label{sec:rq2-results}

\begin{table}[ht]
\small
\caption{RQ2: Verification-tool usage across ADAPT test-generation
runs. The full set of runs includes both F2P and non-F2P cases.}
\label{tab:rq2_trajectory}
\begin{tabularx}{\columnwidth}{@{}Xrr@{}}
\toprule
Trajectory Type & GPT-4o & GPT-5 mini \\
\midrule
Single verification-tool call    & 128 & 252 \\
Repeated verification-tool calls & 292 & 135 \\
Total runs with verification     & 420 & 387 \\
Avg.\ verification calls per run & 5.4 & 1.8 \\
\bottomrule
\end{tabularx}
\end{table}

\autoref{tab:rq2_trajectory} summarizes how the verification tool is
used across generation trajectories. Under GPT-5 mini, $252$ runs
require only a single verification call, while $135$ runs involve
repeated calls, indicating that the agent often relies on intermediate
feedback to revise candidate tests. Under GPT-4o this pattern is
reversed: $292$ runs involve repeated calls versus $128$ with a single
call, showing that the weaker model needs substantially more iterations.
The average number of verification calls per run reflects this gap:
$1.8$ for GPT-5 mini ($696$ calls across $387$ runs) versus $5.4$ for
GPT-4o ($2{,}259$ calls across $420$ runs).

\begin{table}[ht]
\small
\caption{RQ2: Composition of test cases across all verification calls.}
\label{tab:rq2_multitest}
\begin{tabularx}{\columnwidth}{@{}Xrrrr@{}}
\toprule
& \multicolumn{2}{c}{GPT-4o} & \multicolumn{2}{c}{GPT-5 mini} \\
\cmidrule(lr){2-3}\cmidrule(lr){4-5}
Category & Count & \% & Count & \% \\
\midrule
F2P test cases     &   384 &  9.2 &   608 & 23.8 \\
Non-F2P test cases & 3{,}778 & 90.8 & 1{,}943 & 76.2 \\
\addlinespace
Total test cases   & 4{,}162 & 100.0 & 2{,}551 & 100.0 \\
\bottomrule
\end{tabularx}
\end{table}

\autoref{tab:rq2_multitest} shows the composition of all test cases
submitted to the verification tool. GPT-5 mini produces $608$ F2P test
cases out of $2{,}551$ total ($23.8\%$), while GPT-4o produces $384$
out of $4{,}162$ ($9.2\%$). Despite submitting nearly twice as many test
cases overall, GPT-4o yields fewer F2P cases, confirming that the
additional iterations under the weaker model are largely spent on
unsuccessful attempts. Taken together, these results show that the
verification tool is not merely a final checker but an active component
of the generation workflow that supports iterative refinement, and that
its effectiveness scales with model capability.

\begin{findingbox}
\textbf{Finding 2.} The verification tool is used iteratively during generation and is not only a final check.
Of the $6{,}713$ candidate tests submitted across all runs, only $992$ ($14.8\%$) were F2P, so the remaining $5{,}721$ ($85.2\%$) were filtered out by the tool.
The tool was also called repeatedly rather than once in a large share of runs, $292$ of $420$ under GPT-4o and $135$ of $387$ under GPT-5 mini, averaging $5.4$ and $1.8$ calls per run respectively.
\end{findingbox}

\begin{figure}[t]
    \centering
    \includegraphics[width=\columnwidth]{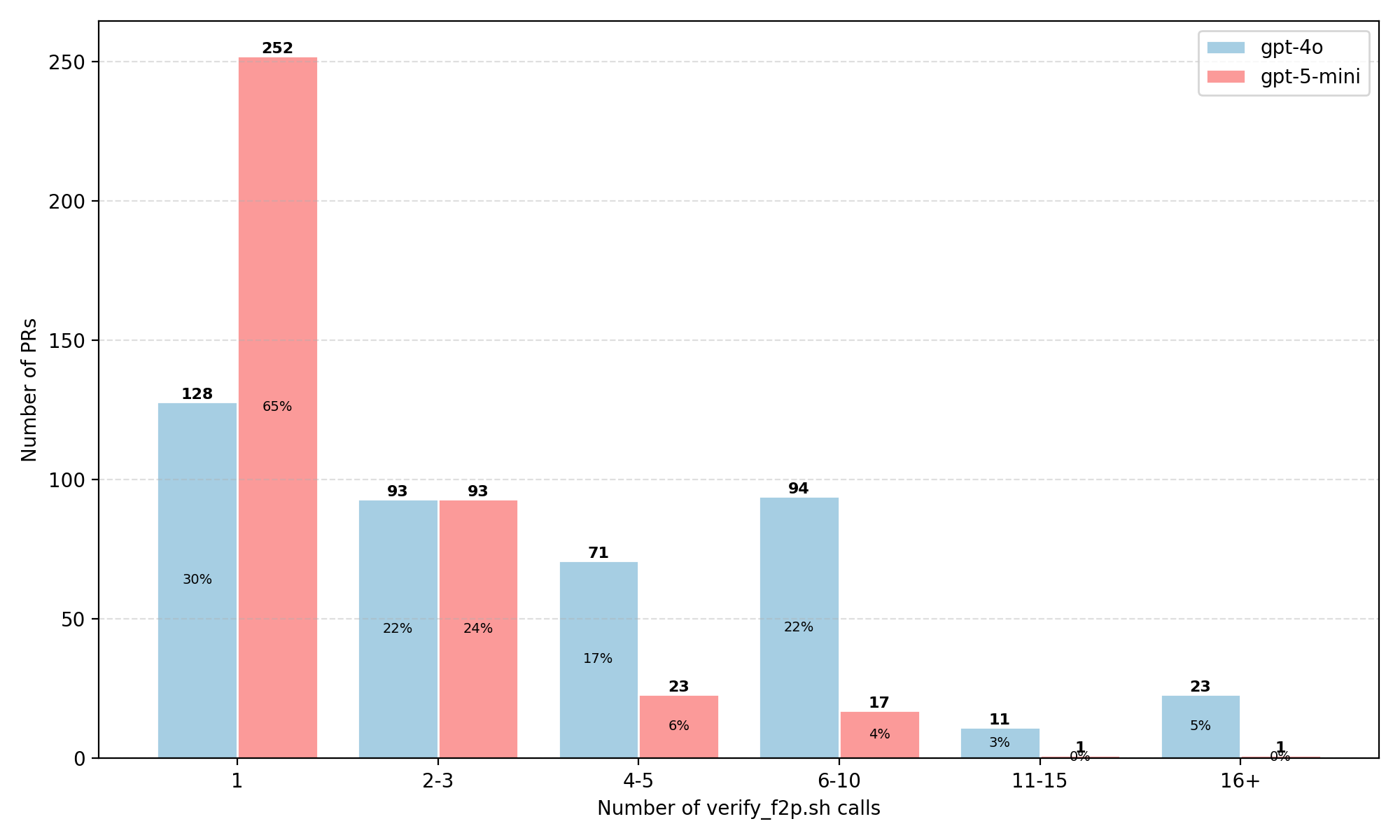}
    \caption{Distribution of F2P verification-tool calls across PRs.}
    \label{fig:bar_chart}
\end{figure}

\subsection{RQ3: \RQthree}
\label{sec:rq3-results}

RQ3 examines whether the valid tests generated by ADAPT meaningfully
target the behavioral change introduced by the PR and are of high
quality. As described in Section~\ref{sec:manual-eval}, four annotators
evaluated a sample of 70 F2P-valid tests under each
configuration, and the results are reported in
Table~\ref{tab:rq3_results}.

\begin{table}[t]
\small
\caption{RQ3: Manual evaluation of 70 F2P-valid tests per
configuration.}
\label{tab:rq3_results}
\begin{tabularx}{\columnwidth}{@{}Xrrr@{}}
\toprule
Question & Mean & AC & MAD \\
\midrule
\multicolumn{4}{@{}l}{\textbf{GPT-4o}} \\
Importance (Q1)         & 4.01 & 0.732 & 0.895 \\
Assertion strength (Q2) & 3.79 & 0.653 & 0.976 \\
Readability (Q3)        & 4.16 & 0.802 & 0.845 \\
Gen.\ $>$ Human (Q4) & 0.66 & 0.371 & 0.348 \\
Human $>$ Gen.\ (Q5) & 0.81 & 0.460 & 0.319 \\
\addlinespace
\multicolumn{4}{@{}l}{\textbf{GPT-5 mini}} \\
Importance (Q1)         & 4.54 & 0.818 & 0.745 \\
Assertion strength (Q2) & 4.35 & 0.843 & 0.688 \\
Readability (Q3)        & 4.49 & 0.884 & 0.629 \\
Gen.\ $>$ Human (Q4) & 0.64 & 0.403 & 0.321 \\
Human $>$ Gen.\ (Q5) & 0.78 & 0.513 & 0.288 \\
\bottomrule
\end{tabularx}
\end{table}

\begin{figure}[t]
    \centering
    \includegraphics[width=\columnwidth]{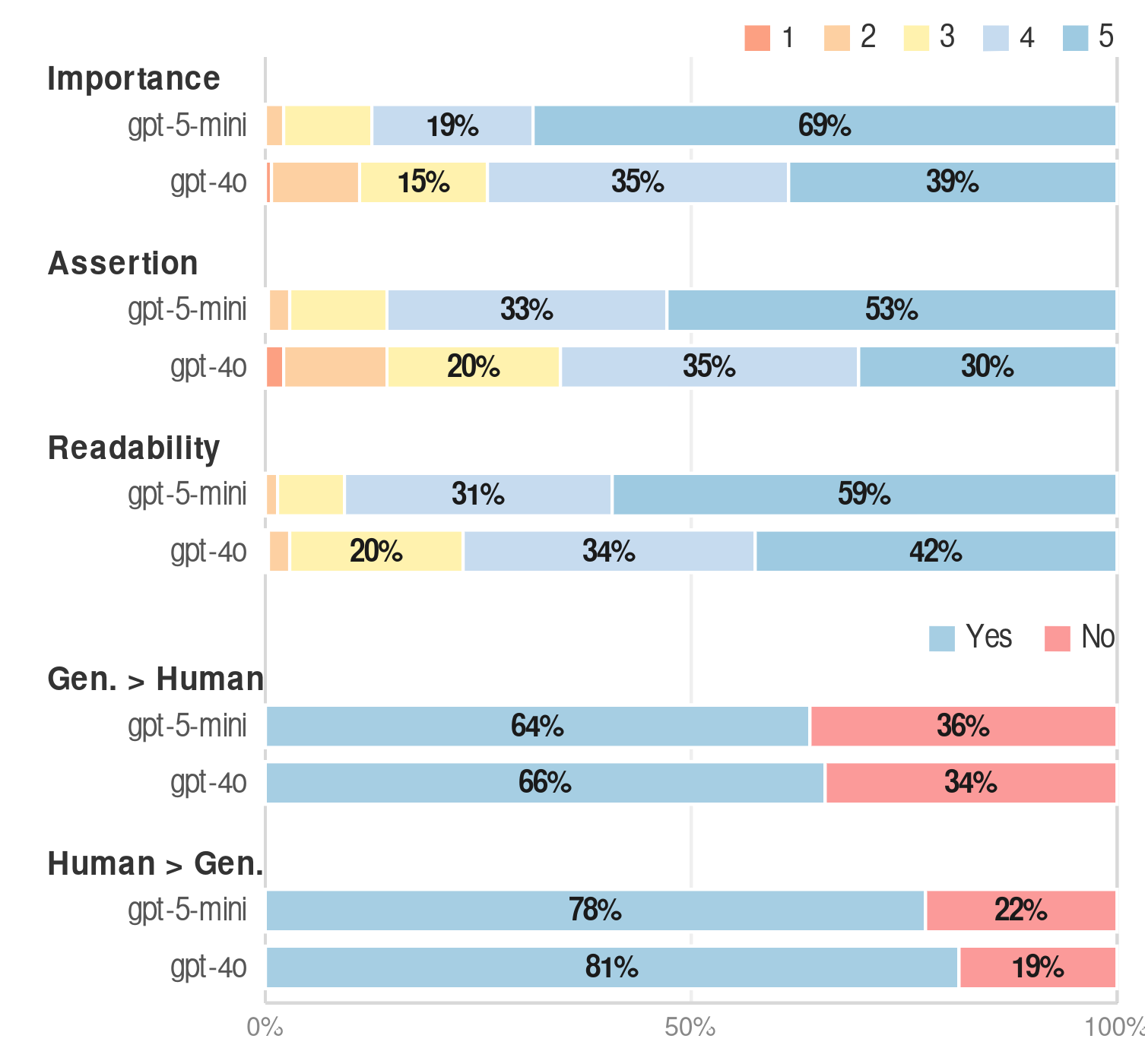}
    \caption{Distribution of the manual-evaluation votes, pooled over
    all four annotators ($280$ votes per dimension).}
    \label{fig:vote_distribution}
\end{figure}

\begin{findingbox}
    \textbf{Finding 3.} F2P-valid tests genuinely target the PR's
    behavioral change. On the sample of $70$ tests, GPT-5 mini achieves
    mean scores of $4.54/5$ for importance, $4.35/5$ for assertion strength,
    and $4.49/5$ for readability; GPT-4o scores $4.01$, $3.79$, and $4.16$.
    The generated tests cover behavior beyond the developer-written test in
    $66\%$ (GPT-4o) and $64\%$ (GPT-5 mini) of cases.
    \end{findingbox}

As shown in \autoref{tab:rq3_results}, the per-test quality dimensions
score highly under both configurations. GPT-5 mini reaches mean scores
of $4.54$ for importance, $4.35$ for assertion strength, and $4.49$ for
readability, each above the corresponding GPT-4o means of $4.01$, $3.79$,
and $4.16$. \autoref{fig:vote_distribution} shows that the underlying
votes are concentrated in the top categories: for GPT-5 mini,
$68.6\%$, $52.9\%$, and $59.3\%$ of votes assign the maximum score of
$5$ to importance, assertion strength, and readability, and $87.5\%$,
$85.8\%$, and $90.7\%$ fall in the $4$--$5$ range. GPT-4o is more
spread but still concentrated in the $4$--$5$ range ($74.0\%$ of
importance votes, $65.4\%$ of assertion-strength votes, and $76.8\%$ of
readability votes fall in these two categories).
Pairwise annotator agreement on these dimensions is substantial, with
Gwet's AC between $0.82$ and $0.88$ across the three dimensions for
GPT-5 mini and between $0.65$ and $0.80$ for GPT-4o, and the mean
absolute differences remain below one point on the five-point scale.
On the comparative questions, the generated tests frequently
cover behavior beyond the developer-written test (Gen.\ $>$ Human:
$0.66$ and $0.64$ for GPT-4o and GPT-5 mini), while the
developer-written test in turn covers behavior the generated test misses
in most cases (Human $>$ Gen.: $0.81$ and $0.78$). Agreement on the
comparative questions is lower than on the per-test dimensions, with
Gwet's AC of $0.460$ and $0.513$ for Human $>$ Gen.\ and $0.371$ and
$0.403$ for Gen.\ $>$ Human. Finally, low ratings are
uncommon: across the three per-test dimensions, GPT-4o received $79$ of
$840$ votes of $2$ or below ($31$ for importance, $40$ for assertion
strength, and $8$ for readability), while GPT-5 mini received $18$ such
votes ($6$ for importance, $8$ for assertion strength, and $4$ for
readability).

\subsection{RQ4: \RQfour}
\label{sec:rq4-results}

\begin{table}[t]
\footnotesize
\setlength{\tabcolsep}{3pt}
\caption{RQ4: Full patch coverage on the 137 PRs shared by ADAPT and ChaCo. Cells report count (rate).}
\label{tab:rq4_coverage}
\begin{tabularx}{\columnwidth}{@{}Xrrrr@{}}
\toprule
Approach & Overall & pandas & SciPy & Qiskit \\
& ($n=137$) & ($n=44$) & ($n=49$) & ($n=44$) \\
\midrule
ADAPT & 115 (83.9\%) & 40 (90.9\%) & 40 (81.6\%) & 35 (79.5\%) \\
ChaCo & 109 (79.6\%) & 35 (79.5\%) & 38 (77.6\%) & 36 (81.8\%) \\
\bottomrule
\end{tabularx}
\end{table}

\autoref{tab:rq4_coverage} compares the full-coverage outcomes of
ADAPT and ChaCo under GPT-5 mini. ADAPT reaches full patch coverage on
$115$ of the $137$ shared PRs ($83.9\%$), compared with $109$ PRs
($79.6\%$) for ChaCo, an absolute improvement of $4.4$ percentage
points. ADAPT performs better on pandas, where it fully covers $40$ PRs
versus ChaCo's $35$, and on SciPy, where it covers $40$ versus $38$.
On Qiskit, ChaCo covers one additional PR ($36$ versus $35$). Across
the complete shared set, both approaches fully cover $93$ PRs, $22$ are
covered only by ADAPT, $16$ only by ChaCo, and $6$ by neither.

\begin{table}[t]
\scriptsize
\setlength{\tabcolsep}{1.5pt}
\caption{RQ4: Resource usage on the 137 PRs shared by ADAPT and ChaCo. Time is reported as mean/median minutes per PR.}
\label{tab:rq4_resources}
\begin{tabularx}{\columnwidth}{@{}Xrrrrrrrrr@{}}
\toprule
Approach & PRs & Input & Cached & Output & Total tok. & Tok./PR & Total\$ & \$/PR & Time/PR \\
\midrule
ADAPT & 137 & 265.7M & 255.4M & 5.4M & 271.1M & 1.98M & 19.82 & 0.145 & 17.4/13.9 \\
ChaCo & 137 & 58.5M & -- & 23.0M & 81.6M & 0.60M & $\sim$60.28 & $\sim$0.44 & 56.7/46.2 \\
\bottomrule
\end{tabularx}
\end{table}

\autoref{tab:rq4_resources} shows that ADAPT uses more tokens than
ChaCo, both in total ($271.1$M versus $81.6$M) and per PR ($1.98$M
versus $0.60$M). Despite this, ADAPT has lower recorded cost:
\$$19.82$ in total and \$$0.145$ per PR, compared with ChaCo's
estimated \$$60.28$ and \$$0.44$ per PR. ADAPT also completes runs more
quickly, averaging $17.4$ minutes per PR with a median of $13.9$
minutes, whereas ChaCo's observed generation spans average $56.7$
minutes with a median of $46.2$ minutes. The ChaCo time values are
calculated over the $135$ PRs for which generation-call timestamps are
available.

\begin{findingbox}
\textbf{Finding 4.} ADAPT outperforms ChaCo in full patch coverage,
covering $115/137$ shared PRs ($83.9\%$) compared with $109/137$
($79.6\%$), an improvement of $4.4$ percentage points. ADAPT achieves
the higher overall rate while requiring less recorded cost and shorter
observed time per PR, although it uses more tokens.
\end{findingbox}

\section{Discussion}

Under the same-model setting, where both ADAPT and BLAST use GPT-4o, ADAPT achieves an F2P rate of 62.0\% compared with BLAST's reported 35.4\%, an absolute improvement of 26.6 percentage points.
This advantage suggests that PR-aware F2P generation benefits from an agentic workflow that can navigate repository context, execute tests, interpret outcomes, and refine candidates from feedback.

A second important implication concerns the role of the dedicated F2P verification mechanism, which the trajectory-level analysis suggests contributes beyond simple final validation.
Repeated verification occurred in $292/420$ GPT-4o runs and $135/387$ GPT-5 mini runs, averaging $5.4$ and $1.8$ calls per run, respectively (\autoref{tab:rq2_trajectory}).
Across both models, only $992$ of $6{,}713$ submitted tests ($14.8\%$) were F2P (\autoref{tab:rq2_multitest}).
Together, these observations indicate that agents frequently use dual-version feedback to refine partially relevant candidates and that GPT-5 mini uses this signal more efficiently.
Because the verifier is present in every ADAPT run, however, these data demonstrate iterative use rather than establishing its causal contribution.

Another implication concerns the value of combining behavioral and structural feedback during generation.
On the 137 PRs shared with ChaCo, ADAPT reaches its full-coverage objective on $83.9\%$ of PRs compared with ChaCo's $79.6\%$, with the advantage concentrated in pandas and SciPy.
This result shows that requiring tests to remain F2P while using uncovered changed lines as refinement feedback does not prevent ADAPT from attaining strong structural coverage~\cite{dipongkor2026coverage}.
The comparison also exposes a resource trade-off: ADAPT uses more tokens per PR, but its runs have lower recorded cost and shorter observed generation time than ChaCo.

\paragraph{Future Work.}
Future work should evaluate ADAPT across broader model families and capacity levels, including Claude, Gemini, and open-weight models.
Industrial studies on test-less PRs should also examine the usefulness of generated tests and their integration into development workflows through practitioner feedback.

\section{Threats to Validity}

\textbf{Internal Validity.}
A threat to internal validity arises from the use of an autonomous
LLM-based agent, whose behavior may vary due to model
stochasticity and the specific action trajectory followed during
each run. In addition, the observed performance can be influenced
by implementation choices, such as prompt design, interaction
budget, and how the repository context is presented to the agent~\cite{baltes2026evidence}.
We mitigate these risks by using a standardized containerized
execution environment and by applying the same benchmark
setting and F2P-based evaluation protocol used in BLAST.
Single-run verification may misclassify flaky tests~\cite{eck2019flaky}.
Another potential threat is data leakage, since the benchmark is
constructed from publicly available repositories and pull requests
that may have been seen during pretraining~\cite{openai2026sweverified}. This threat cannot be
fully eliminated and is a general limitation of evaluations based on
publicly available software artifacts~\cite{djire2026memorized,molinelli2025oracles}. Importantly, however, this issue is not unique to ADAPT and also applies to BLAST, since in our same-model comparison both approaches are evaluated on the same 426 PR instances, drawn from the same public repositories, and use the same underlying LLM, GPT-4o. Therefore, although the possibility of prior exposure cannot be ruled out, using the same dataset, model, and evaluation protocol preserves the fairness of the comparison. Furthermore, our trajectory analysis provides indirect evidence against the concern that the agent merely reproduces memorized solutions: across runs, the agent performs repository inspection and issues multiple F2P verification calls before arriving at a valid test (Section~\ref{sec:rq2-results}). Such iterative inspection and refinement would be largely unnecessary if the agent were simply recalling the developer-written tests from pretraining, suggesting that ADAPT derives its tests from the actual repository state rather than from memorized artifacts.

\textbf{Construct Validity.}
A threat to construct validity concerns whether the F2P criterion fully
reflects test quality, since a single metric may not capture every
desirable property of a generated test. For instance, some generated
tests that are not F2P, and thus not directly aimed at the PR change,
may still be valuable in a broader testing sense. However, the goal of
this study is specifically to generate tests that capture the
behavioral change introduced by a PR, and F2P is the most widely used
criterion for this purpose. We therefore deliberately adopt it, also
because it is the same criterion used by the prior state-of-the-art
study, which enables a fair and methodologically consistent comparison
with the baseline under the same benchmark and experimental setting.
F2P measures change sensitivity, not PR correctness or bug-finding ability~\cite{konstantinou2026risk,liu2023realistic}.

\textbf{External Validity.}
One of the threats to external validity is the generalizability of the results beyond the evaluated benchmark. ADAPT is assessed on a subset of 426 PRs from SWE-Bench Verified, consisting of real-world Python instances that span 12 open-source repositories. Although this benchmark is appropriate for a controlled comparison with prior work, the findings may not generalize to projects written in other programming languages~\cite{lequang2026xrepotest}, or to PRs whose changes are substantially larger, more cross-cutting, or less testable through unit tests alone. Future studies on additional datasets, repositories, languages, and industrial settings would help assess the generalizability of the proposed approach.

\section{Conclusion}
In this work, we presented ADAPT, an agentic diff-aware approach for PR-level test generation that combines autonomous repository interaction with F2P verification. Built on OpenHands, ADAPT augments a general-purpose software engineering agent with PR-specific context and a dedicated F2P verification tool that guides iterative refinement toward change-focused tests capturing the behavioral intent of a patch.

On the same subset of SWE-bench Verified, ADAPT outperforms BLAST under the same F2P evaluation, achieving an F2P rate of 62.0\% versus 35.4\% under the same model (GPT-4o), rising to 89.9\% with GPT-5 mini.
Our trajectory-level analysis shows the verification tool contributes beyond simple final checking, as many successful instances require multiple rounds of dual-version feedback before an F2P test is obtained.
On the 137 PRs shared with ChaCo, ADAPT also reaches full patch coverage on 83.9\% of PRs compared with 79.6\% for ChaCo, while requiring lower recorded cost and shorter observed time per PR.

More broadly, the results suggest that PR-aware test generation should be viewed not only as a prompting problem but as an interactive reasoning and refinement task. Because a large fraction of merged PRs are submitted without accompanying tests, ADAPT offers a practical path toward automatically generating change-focused tests for such PRs, helping close this gap and protect the introduced changes against future regressions. We hope this work motivates further research on agentic testing systems that combine context navigation, execution-based feedback, and patch-aware behavioral testing.

\bibliographystyle{IEEEtran}
\bibliography{references}

\end{document}